\documentclass{elsart}

\usepackage{amsmath}
\usepackage{amssymb}
\usepackage{graphicx}
\usepackage{dcolumn}
\usepackage{bm}

\def\be{\begin{eqnarray}}
\def\ee{\end{eqnarray}}
\def\ba{\begin{array}}
\def\ea{\end{array}}
\def\nn{\nonumber}
\begin{document}
\begin{frontmatter}
\title{The spin-split incompressible edge states within empirical
Hartree approximation at intermediately large Hall samples}
\author{A. Siddiki}
\ead{siddiki@lmu.de}
\address{Center for NanoScience and Dept. for Physics, Ludwigs-Maximillians University,
D-80539 Munich, Germany}
\begin{abstract}
A self-consistent Thomas-Fermi-Poisson based calculation scheme is
used to achieve spin resolved incompressible strips (ISs). The
effect of exchange and correlation is incorporated by an
empirically induced $g$ factor. A local version of the Ohm's law
describes the imposed fixed current, where the discrepancies of
this model are resolved by a relevant spatial averaging process.
The longitudinal resistance is obtained as a function of the
perpendicular (strong) magnetic field at filling factor one and
two plateaus. Interrelation between the ISs and the longitudinal
zeros is explicitly shown.
\end{abstract}
\end{frontmatter}
%
Devising spin degree of freedom as a coherent state to carry
information became an essential phenomena in today's quantum
information processing research, also in the integer quantized
Hall effect~\cite{vKlitzing80:494} (IQHE) regime. However, a
consistent picture describing the current distribution within
these samples is still under debate. Recent attempts to calculate
these quantities, can be categorized under two essential titles:
(i) A semi-classical solution of the Poisson and Schr\"odinger
equations to obtain electrostatic
quantities~\cite{Oh97:13519,Siddiki03:125315}, i.e. a Hartree type
(self-consistent) Thomas-Fermi(-Poisson) approximation (SCTFPA),
in which the imposed current is treated within a local version of
the Ohm's law~\cite{Guven03:115327,siddiki2004} (LOL), (ii) A full
quantum mechanical solution of the Hartree-Fock Hamiltonian,
either using direct diagonalization\cite{Romer07} or local spin
density approximation (LSDA) plus density functional theory
(DFT)~\cite{Igor06:075320}. The latter two approximations,
although much powerful to describe the electrostatics, lack
presenting the global longitudinal or Hall resistances, since the
direct diagonalization model is essentially focused on describing
the local compressibility or conductivity and the second one uses
the commonly used Landauer-B\"uttiker type conductance phenomena.
On the other hand, the simpler approach, i.e. SCTFPA+LOL, was able
to provide information of the local current distribution and the
global resistances as a function of magnetic field considering
spinless electrons~\cite{Siddiki:ijmp}. Here, we extend our
previous works~\cite{siddiki2004,Siddiki:ijmp} to include the
effect of spin, i.e. Zeeman splitting, by considering an
experimentally induced $g$ factor~\cite{Khrapai:05}. The simplest,
but not the most trivial, treatment of Zeeman splitting,
conductivity model and exchange-correlation effects is used to
describe the mentioned quantities. We, essentially incorporate
Zeeman effect by introducing a gap in the energy dispersion, then
calculate the electron and the current densities together with the
electrostatic and electrochemical potentials as a function of the
lateral coordinate. Moreover, the longitudinal resistance
$(R_{xx})$ is calculated for varying magnetic field.
\begin{figure}
{\centering
\includegraphics[width=1.\linewidth]{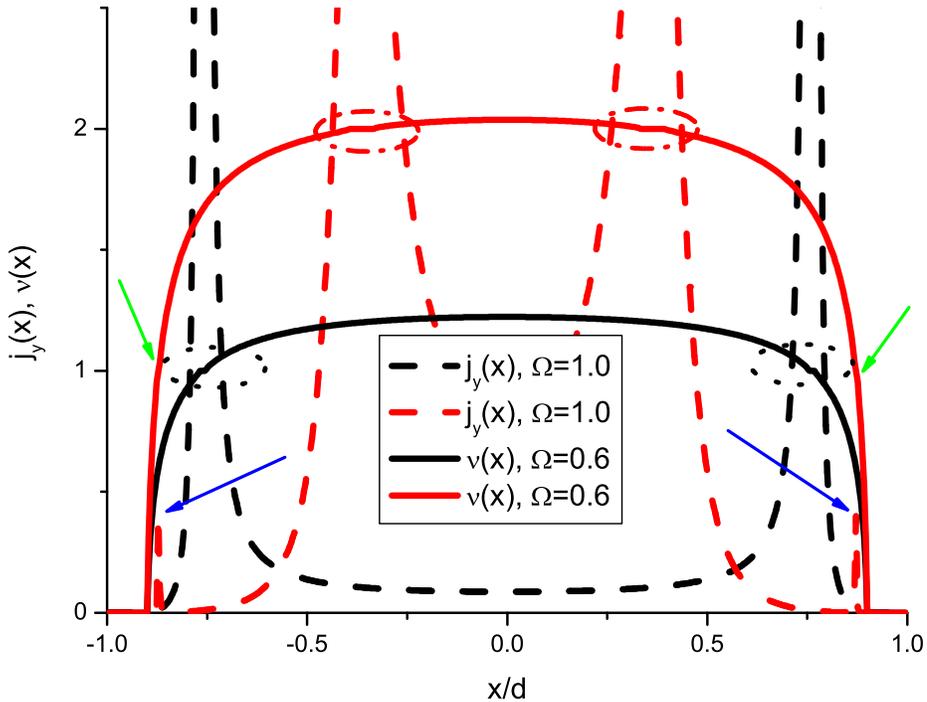}
%
\caption{ \label{fig:fig1} The local filling factors (solid lines)
calculated at two representative magnetic field values,
$\Omega=\hbar\omega_c/E_0$, where $\hbar\omega_c=\hbar eB/m$ is
the cyclotron energy and $E_0$ is the pinch-off energy given by
$2\pi e^2n_0/\kappa$, with the fixed donor number density
($n_0=4\times10^{11}$ cm$^{-2}$). Corresponding current densities
(broken lines) a small current ($I=U_H/\Omega=0.01$, see
Ref.\cite{Guven03:115327} for the definition) is driven in $y$
direction, which is still in the linear response regime. The
calculations are done at the default temperature $k_BT/E_0=0.001$
for a sample width $2d=6.3$ $\mu$m.}}
\end{figure}
In our calculations we assume a translational invariant sample in
$y$ direction, where an electron channel is formed in the interval
$-b<x<b$, with $b$ being the depletion length. The geometry under
consideration is the so-called floating gate
structure~\cite{Guven03:115327}. Given the boundary conditions and
external potentials one can solve the Poisson equation, starting
from the zero temperature, zero magnetic field solution. The basic
ingredients of the SCTFPA are the iterative solution of the
electron density, \be n_{el}(x)=\int dE D(E)f(E-V(x)),\ee where
$f(\alpha)$ is the Fermi function, $D(E)$ a relevant density of
states, and the total potential energy, \be
V(x)=&&V_{ext}(x)+(2e^2/\kappa)\int_{-d}^{d}n_{el}(x')K(x,x')dx'
\nn
\\&& + g^*\sigma \mu_B B .\ee Here, the first term stands for the
external potential composed of gates and donors. In the second
(Hartree) term $\kappa$ is the dielectric constant of the material
and $K(x,x')$ is the solution of the Poisson equation preserving
the boundary conditions $V(-d)=V(d)=0$, where $2d$ is the sample
width. The third term is the Zeeman energy with effective $g$
factor ($g^*=5.2$ from experimental and theoretical
findings~\cite{Khrapai:05} and the references therein),
$\sigma=\pm1/2$ spin polarization and $\mu_B$ the effective Bohr
magneton considering GaAs/AlGaAs heterostructures. The well
established wisdom is that, the exchange-correlation effects
enhance the Zeeman splitting, however, direct numerical
calculations of these potentials without taking into account
screening lead to discrepancies in explaining the experimental
data. In this work we take the experimental value of the $g$
factor without dealing with the difficulties that arise from
explicit calculation of the effects of exchange and
correlations~\cite{Romer07,Igor06:075320}.

The imposed current densities are described by a LOL, by making
use of the translation invariance and equation of continuity,
namely \be E_y(x)\equiv E_y^0,\quad {\rm and} \quad j_x(x)=0, \ee
and \be j_y(x)=E_y^0/\rho_{xx}(x),\quad
E_x(x)=\frac{\rho_{xy}(x)}{\rho_{xx}(x)}E_y^0,\ee where
$\rho_{xx}(x)$ and $\rho_{xy}(x)$ are the diagonal (longitudinal)
and off-diagonal (Hall) components of (local) resistivity tensor,
$\hat{\rho(x)}=[\hat{\sigma(x)}]^{-1}$, respectively. Hence, given
the local conductivity tensor elements one can obtain the current
density in $x$ direction and thereby the global resistances. We
choose a simple (generalized) description of the conductivity
model, namely the Hall component is $\sigma_{xy}=(e^2/h)\nu(x)$,
where $\nu(x)$ is the local filling factor and the longitudinal
component is  $\sigma_{xx}=$ \be
 \sigma_{xy}[\epsilon+(1-\nu(x))^2/4], & \quad 0<\nu(x)<1.5 \nn \\
  \sigma_{xy}[\epsilon+(2-\nu(x))^2/4], & \quad 1.5\le \nu(x)<2.5
,\ee where $\epsilon$ is a cut-off parameter defining the accuracy
of the numerics ($\sim 2\times 10^{-6}$). The above scheme,
enclosures the simplest set of assumptions to provide a
qualitative understanding of the current distribution and the
calculation of the global resistances and can be improved in many
aspects, however, as a first attempt grasps most of the essential
physics.

\begin{figure}
{\centering
\includegraphics[width=1.\linewidth]{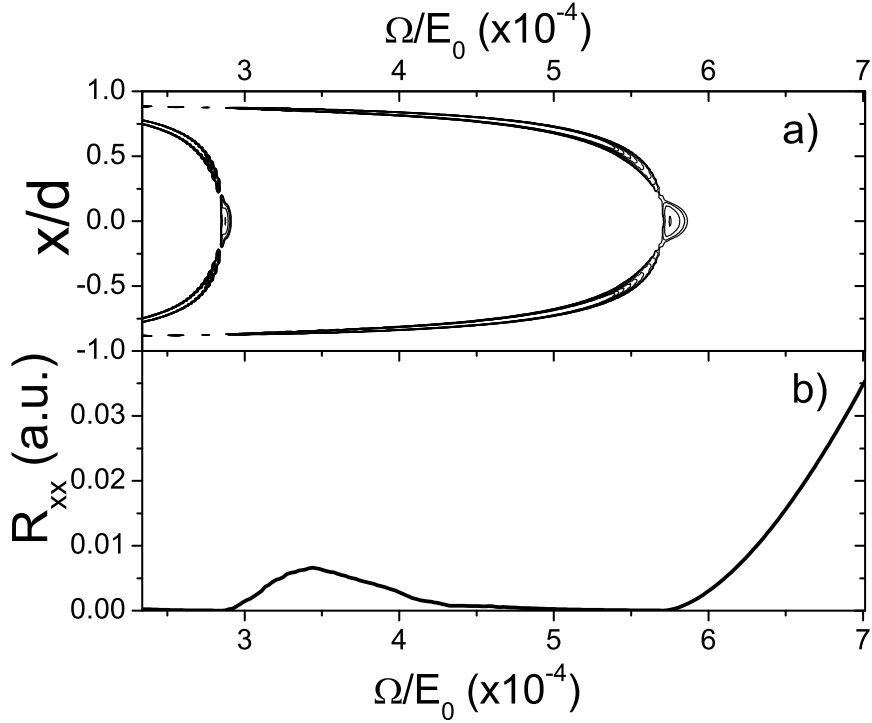}
%
\caption{ \label{fig:fig2}(a) The numerically calculated current
distribution as a function of external magnetic field, lines
present the equi-current contours. (b) $R_{xx}$ against magnetic
field, obtained at default temperature.}}
\end{figure}
Fig.~\ref{fig:fig1} depicts the calculated local filling factors
versus the lateral coordinate, when the $\nu(0)<2$ (black solid
line) and $2<\nu(0)<3$ (red solid line). The ellipses encircle the
region where the local filling factor is one (dotted) or two
(dash-dotted). It is seen apparently that, the current (broken
lines) is essentially confined into a region where an IS exists.
In the case of $\nu(0)<2$, most of the current is flowing either
inside the IS or in the very close vicinity, whereas for
$2<\nu(0)<3$ beside the main two peaks near filling factor two, we
observe that there exists two other peaks on both sides coinciding
with the positions of filling factor one regions (highlighted by
arrows). This essentially indicates that although most of the
current is carried by the innermost IS, if an external current is
imposed there is a possibility for electrons to flow from these
very narrow local regions, due to a local minimum of the
longitudinal conductivity. This finding coincides with the very
recent model by I. Neder~\cite{Izhar07:ngauss}, considering a
non-Gaussian noise to explain the visibility oscillations
performed at the Mach-Zehnder interference experiments. In
Fig.~\ref{fig:fig2} we show the local current intensity (a) as a
function of magnetic field together with the longitudinal
resistance (b). The evolution of the reminiscence IS with
$\nu(x)=1$ can also be observed in the left side of the current
distribution, however, the longitudinal resistance vanishes within
the numerical accuracy in this interval. If the ISs become large
than the magnetic length or the Fermi wave length, the system
exhibits finite $R_{xx}$ and the current is distributed all over
the sample, otherwise $R_{xx}=0$. The bubble-like features seen at
the centers of the sample is simply due to the conductivity model
we consider.

In summary, we have calculated the spatial distribution of the
current carrying, spin-split ISs within a Hartree type
approximation incorporating Zeeman splitting by an experimentally
induced effective $g$ factor. We have shown that, the formation of
(spin resolved) incompressible strips dominate the current
distribution and thereby defines the magnetic field interval where
one would observe the IQHE.

The author acknowledges, SFB631 and DIP for financial support and
A. Ulubay S. for critical reading of the manuscript.

\end{document}